8-26-2021

# A Study and Analysis of Manuscript Publications in the Open Access Journals

Subaveerapandiyan A
*Regional Institute of Education Mysore*, subaveerapandiyan@gmail.com

Supriya Pradhan
*INFLIBNET Centre*, supriyapradhan.deep@live.com

R Nandhakumar
*Regional Institute of Education, Mysuru*, nandha_7025@yahoo.com

# A Study and Analysis of Manuscript Publications in the Open Access Journals

**Subaveerapandiyan A**

Professional Assistant

Regional Institute of Education, Mysore, India

E-mail: subaveerapandiyan@gmail.com

**Supriya Pradhan**

Scientific Technical Assistant (LS)

INFLIBNET Centre, India

E-mail: supriyapradhan.deep@live.com

**R. Nandhakumar**

Assistant Professor (Contractual), Department of Education,

Regional Institute of Education, Mysuru, India

Mobile: 9842408251, E-mail ID: nandha_7025@yahoo.com

**Abstract**

*The purpose of this study is to analyze the research article publishing with special reference to preparing to publish and peer-reviewing. Peer reviewing is the process required for standardizing any publications. Manuscript writing is an art. Though it appears to be simple there is a lot of effort required. Peer-reviewing is the process that eliminates articles that do not meet the standard of the journals and the scope of the journals. The study investigated authors' views on manuscript submissions to the publishing process. There are 375 samples selected for this study who have experienced publishing journals listed in refereed journals. For the selection of the sample 50 ScimagoJR Library and Information Science open access journals between 2019-2021 are verified by the authors.*

## 1. Introduction

Nowadays, almost all library professionals are interested in publishing articles in library and information science journals, reading, writing, reviewing, or suggesting topics. At the same time, publishing has become a tough task for professionals owing to particular academic and professional necessities. Authors and editors have different perspectives when publishing in reputed journals.

Peer-reviewed journals are called refereed journals. Peer reviewing is majorly classified into four categories namely: open peer-review, single-blind, double-blind, and transparent peer-review. The peer reviewer's job is to evaluate scholarly articles, validating the data, and checking the quality of the content. The peer-reviewing process in the 21st Century is gradual but steady in the race. Journals should meet high standards in their publications. Reviewers must be concerned with the fast process because many countries' educational institutes make the Scopus and scholarly publications compulsory for completing the PhD. Peer reviewers are experts and they have been assigned to evaluate the enormous papers effectively, efficiently and give the updates reviewing, revising, rejecting, and ready to publish, accept or reject status as earlier.

### 1.2. ScimagoJR

SCImago Journal Rank (SJR) provides the journal metrics, and it furnishes the details of Scopus indexed journals ranking, subject ranking and country ranking. It gives a search facilities title, ISSN or publisher to the users. As well as it provides the advanced searching and filter details they are divided: 27 subject areas, 313 subject categories, 5000+ international publishers from 239 countries, all types of sources (journals, books, conference and proceedings, and trade journals), years covered 2000 to 2020, and other filter options are i) Only open access journals ii) SciELO (Scientific Electronic Library Online) journals and iii) WoS (Web of Science) journals. Scimago developed by Scimago lab. Total open access journals in 2020 listed 6885 and Library and information science-based open access journals are 65.

### 2. Review of Literature

Khalifa & Ahmed (2021) conducted a research study on orthopedic related journal paper publications peer-review process time during the Covid-19. For this study they used the PubMed database to use the keyword orthopedic and filtered the publications from 2019 December 1 to 2020 august 1 after that they downloaded 231 articles. In these 231 articles, they tried to find author article submission time to publication time. Their study result found that the peer-review process took less than 30 days.

Mavrogenis et al. (2020) discussed in their article peer reviewers are not interested in communicating with authors and editors during the reviewing time. But double peer blind review articles are not applicable. They explain a good review process. Critical denunciation has to give the reviewer to the editor before rejecting or revising the paper. They have to explain which part of the paper research made the mistakes such as novelty in writing, significance, objectives, method, techniques, analysis, or scope because it will be helpful to the author to develop their manuscript.

Ali & Watson (2016) discussed the importance of the peer-reviewing process, types, roles of reviewers and criteria. In their paper, they discussed various types of peer reviews, advantages and disadvantages. They are single-blind reviews; it means reviewers know about the author details and affiliation but scholars don't know who did the review. Double-blind reviews: it means researcher and reviewers identity is secret only the editors know. Open peer reviews: author and reviewer know both identities, sometimes authors can choose their reviewer.

## 3. Objectives of the Study

- ➢ To know the scholar's viewpoint on the article accepting time
- ➢ To understand the researcher's expectations of journal publishing duration
- ➢ To evaluate editorial supports, they get it or not
- ➢ To find out the authors standard citing format
- ➢ To analyze the manuscript published in the open access

## 4. Methodology

This study adopted a survey method and simple random sampling used for this study. The author's email IDs were collected from their published research articles SCImago from 2019 to 2021. All the participants were authors of 50 open access Scopus indexed journals in the Library and Information Science field in the year 2021. The open-ended questionnaires were prepared, and an online survey was conducted. The survey has two parts of questionnaires namely: Part-I demographic details and Part II Manuscript process and the authors' perspectives. Participants belong to various countries in Africa, Asia, Australia/Oceania, Europe, North America, and South America. We did not give any compensation for this study to the authors, and it's completely voluntary based.

## 5. Data Analysis and Interpretation

**Table 1: Socio-demographic Details**

| Socio-Demographic Details | Item | Frequency | Percentage |
|---|---|---|---|
| **Educational qualifications** | PhD | 228 | 60.8 |
| | M.Phil. | 15 | 4 |
| | Post Graduate | 114 | 30.4 |
| | Graduate | 18 | 4.8 |
| **Continents** | Asia | 204 | 54.4 |
| | Africa | 27 | 7.2 |
| | Europe | 78 | 20.8 |
| | North America | 36 | 9.6 |
| | South America | 27 | 7.2 |
| | Australia/Oceania | 3 | 0.8 |
| **Number of Publications** | 1 to 5 | 138 | 36.8 |
| | 6 to 10 | 60 | 16 |
| | 11 to 20 | 54 | 14.4 |
| | 21 to 30 | 24 | 6.4 |
| | 31 to 50 | 45 | 12 |
| | Above 50 | 54 | 14.4 |

Above table 1 shows the socio-demographic details of the respondents, which reveals that the educational qualification of the respondents ranges from graduates to Ph.D. level whereas the majority of them were Post Graduates (30.4%) and PhD Holder (60.8%). Major respondents belong to the regions like Asia and Europe.

**Table 2: Experience with manuscript submission**

| Duration | Number of Publications | Acceptance | | | | | Rejection After One year % |
|---|---|---|---|---|---|---|---|
| | | Immediately % | In < month % | in <3 months % | in <6 months % | in < 1 year % | |
| Average duration for manuscript acceptance | Less than 5 | 4.3 | 13 | 32.6 | 19.6 | 30.5 | 29.4 |
| | 6-20 | 2.7 | 15.8 | 21 | 31.6 | 28.9 | 27.5 |
| | 21-50 | 0 | 13 | 21.7 | 30.5 | 34.8 | 29.4 |
| | More than 50 | 11.1 | 11.1 | 33.4 | 27.8 | 16.6 | 13.7 |
| Average duration for manuscript publication | Less than 5 | 0 | 15.2 | 23.9 | 23.9 | 37 | NA |
| | 6-20 | 2.6 | 7.9 | 23.7 | 15.8 | 50 | NA |
| | 21-50 | 0 | 0 | 26.1 | 21.7 | 52.2 | NA |
| | More than 50 | 0 | 5.6 | 11.1 | 27.8 | 55.5 | NA |

Based on the general opinion of the authors as given in the Table-2, it is inferred that the experienced authors with more than 50 publications are getting manuscript acceptance quickly. However, they will wait for almost 6 months to 1 year for publication. Authors with lesser experience are getting acceptance depending upon their performance, plagiarism etc., There is no consistency in acceptance and publication. Thus the experience is the key for manuscript acceptance and publication.

**Table 3: Authors General Perspectives on Time Factor**

| Time required | < month | < 3 months | < 6 months | < 1 year | > 1 year |
|---|---|---|---|---|---|
| Reviewing | 138 (36.8%) | 180 (48%) | 42 (11.2%) | 12 (3.2%) | 3 (0.8%) |
| Revising | 174 (46.4%) | 159 (42.4%) | 33 (8.8%) | 6 (1.6%) | 3 (0.8%) |
| Accepting | 108 (28.8%) | 135 (36%) | 108 (28.8%) | 21 (5.6%) | 3 (0.8%) |
| Publishing | 51 (13.6%) | 126 (33.6%) | 126 (33.6%) | 60 (16%) | 12 (3.2%) |

**Figure 1: Authors General Perspectives on Time Factor**

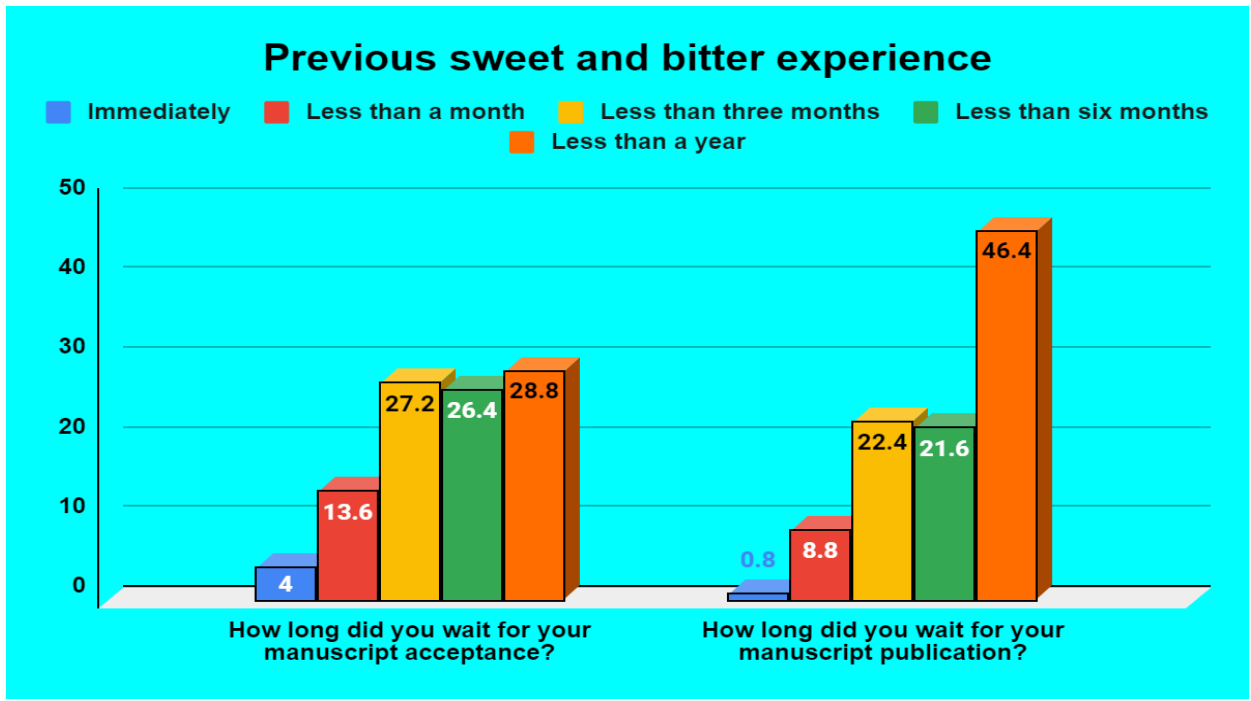

From the table 3 and figure 1 it is clear that time duration within 3 months is optimum and researchers are generally expecting the acceptance during this period. For the reviewing period, 48% of respondents expected acceptance in less than three months. To accept the manuscript time, 36% of respondents expects less than three months and the entire manuscript publishing process in the journal 33.6%, most respondents expects six months and/or less than three months.

**Table 4: How do you feel delaying the process of the manuscript reviewing?**

| How do you feel delaying? the process of the manuscript reviewing? | Frequency | Percentage (n=375) |
|---|---|---|
| Feeling insecure because of someone possible to copy my title | 60 | 16 |
| Feeling insecure because of someone likely to copy my concept | 105 | 28 |
| Feeling insecure because of someone possible to publish an earlier same study | 162 | 43.2 |
| Feeling insecure because of topic will be outdated | 237 | 63.2 |
| Other | 30 | 8 |

Table 4 reported about the delay in reviewing and author's mentality. 63.2% were feeling insecure because topic may be outdated, 43.2 were feeling insecure because of someone likely to publish the same study.

**Table 5: Does a high impacted journal take more time to accept a manuscript?**

| Does the high impacted journal take more time to accept a manuscript? | Frequency | Percentage |
|---|---|---|
| Yes | 210 | 56 |
| No | 78 | 20.8 |
| I do not know | 87 | 23.2 |

Table 5 implies the fact that high impacted journals will take more time to accept the manuscript. 56% of respondents say yes and 20.8% say no and 23% unanswered.

**Table 6: How did you feel your paper rejected after one year of reviewing?**

| Authors feelings of article rejected after 1 year of reviewing | Frequency | Percentage (n=375) |
|---|---|---|
| Angering | 84 | 22.4 |
| Discouraging | 183 | 48.8 |
| Disgusting | 72 | 19.2 |
| Encouraging | 24 | 6.4 |
| Fearing | 6 | 1.6 |
| Other | 66 | 17.6 |

**Figure 2: How do you feel editors do not reply to your queries?**

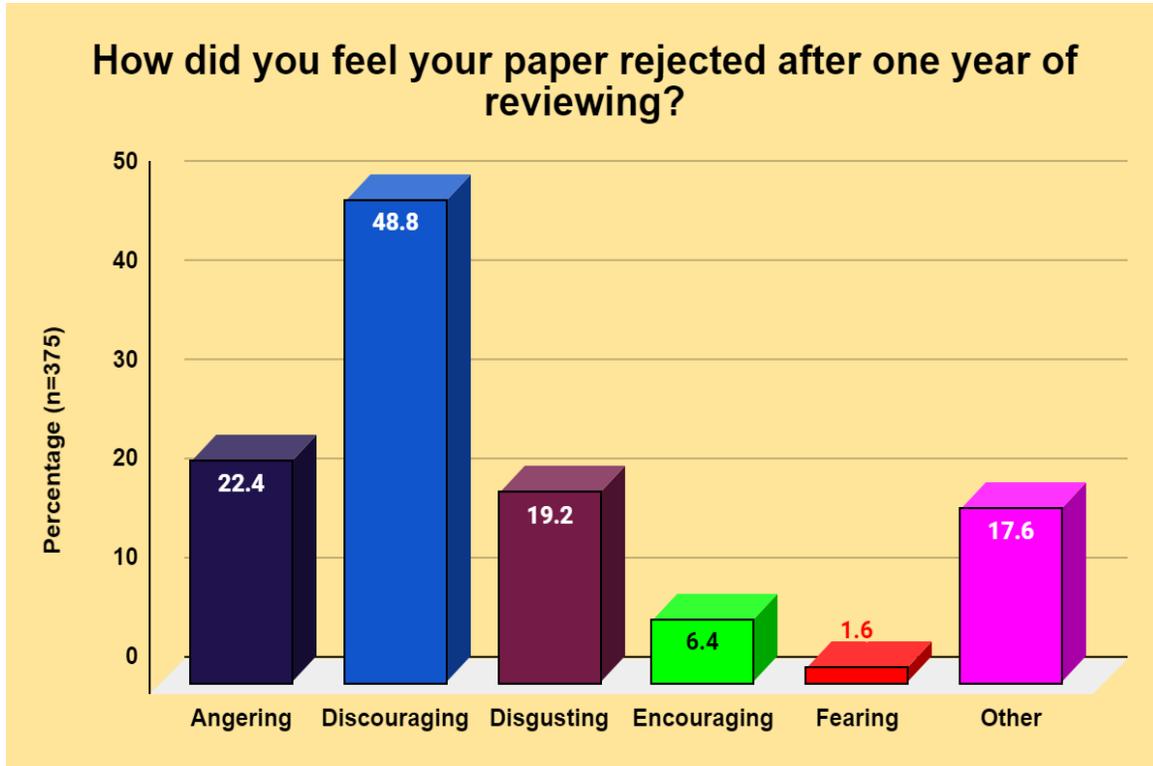

From the Table 6 and figure 2 it is evident that the majority of (48.8%) author got discouraged when the manuscript is rejected after more than one-year review process. 22.4: % got angered while 19.2% got disgusted. Surprisingly 1.6% are getting fear due to the rejection and they may be the inexperienced authors.

**Table 7: Have you received editorial support from the publishers?**

| Have you received editorial support from the publishers? | Frequency | Percentage |
|---|---|---|
| Yes | 219 | 58.4 |
| No | 156 | 41.6 |

Editorial support is the boon for any publication. The above table 7 reported that authors received support from publishers. 58.4% got support from the publisher However, 41.6% did not get any support from the publisher.

**Table 8: Journals are having various citing formats; which format is suitable for all kinds of journals (Opinion)**

| Citing format for all kind of journals (Opinion) | Frequency | Percentage |
|---|---|---|
| APA | 291 | 77.6 |
| MLA | 18 | 4.8 |
| Chicago | 30 | 8 |
| Other | 36 | 9.6 |

Table 8 shows the impact of various referencing formats. APA format is popular among the authors who have published in the peer review journals.

**Discussion**

A survey was undertaken to compile all data on manuscript publications in the existing peer reviewed journals. Data from 375 independent studies were included in this research. All data regarding participants' roles (experienced or novice authors), the methodological approach taken, the type of manuscript, the variables analyzed, and the organizational matters are included in the article. The main goal of the study was to determine the researcher's view on manuscript publication in the peer reviewed journals. Statistical analyses (percentage and percentile) were carried out for the variables. Editorial support is recommended based on the high difference of effect sizes shown in this study. A shorter duration for acceptance (less than 8 weeks), briefer comments, and three or fewer reviews are also suggested. Also, the work will be more suitable to be included in high-indexed journals, reviews, and meta-analyses, facilitating a wider and more rigorous study in the field by future researchers.

Finally, the study recommends

a) Worldwide uniformity in referencing format
b) Editorial supports with minimal cost or free of cost
c) Timely responses through email from the respective authority
d) Preprints for avoiding plagiarism
e) Providing permanent identifiers such as DOI
f) Proper explanation for rejecting and revising the article
g) Minimal Article-processing charges (APC)
h) Universally one open-source plagiarism tool/software

## 6. Conclusion

Manuscript publication has been found to improve academic credentials of people who are involved actively in research works in order to develop their careers. They adhere to basic research principles and this will empower researchers. Interventions among peer reviewers, less duration of reviewing (<2 months), publications (<5 months), and cost effectiveness will encourage authors to maximize research outcomes. Although this study showed greater advantages for experienced researchers, novice researchers are getting discouraged to some extent. At the same time, peer-reviewing has also been beneficial for novice researchers when it gives comprehensive feedback and continuous support to them. The approach taken, the number of participants, or the type of manuscript should not significantly alter the academic outcomes. Although this study suggests implementing peer reviewing in short duration, practitioners should also find academic benefits in any scenario, as academic gains have been documented overall under any condition.